\documentclass[10pt]{article}

\usepackage[utf8]{inputenc}
\usepackage[T1]{fontenc}
\usepackage[english]{babel}
\usepackage[toc,page]{appendix}
\usepackage[normalem]{ulem}

\usepackage{amsmath,amsfonts,amssymb,amsthm}
\usepackage{graphicx}
\usepackage{float}
\usepackage{hyperref}
\usepackage{dsfont}
\usepackage{subcaption}
\usepackage[format=plain,font=small,textfont=it]{caption}
\usepackage[dvipsnames]{xcolor}
\usepackage{multirow} 
\usepackage{bookmark}
\usepackage{mathrsfs}
\usepackage{enumitem}

\graphicspath{{img/}}

\usepackage[numbers,sort&compress]{natbib}

\newcommand{\ket}[1]{|#1\rangle}
\newcommand{\bra}[1]{\langle #1|}
\newcommand{\sket}[1]{|#1]}
\newcommand{\sbra}[1]{[ #1|}

\newcommand{\braket}[2]{\langle #1 \vert #2 \rangle}

\newcommand{\R}{\mathbb{R}}
\newcommand{\C}{\mathbb{C}}

\renewcommand{\d}{\mathrm{d}}

\newcommand{\SixJ}[2]{\left\lbrace \begin{array}{ccc} #1 \\ #2 \end{array}\right\rbrace}

\usepackage{authblk}

\begin{document}

\title{Spinfoam tunneling of quantum geometries in angle variables}

\author[1]{Pietro Don\`a\thanks{pietro.dona@cpt.univ-mrs.fr}}
\author[2,4]{Hal M. Haggard\thanks{hhaggard@bard.edu}}
\author[1,3,4]{Carlo Rovelli\thanks{rovelli@cpt.univ-mrs.fr}}
\author[1]{Gowrisankar Sreeram\thanks{gowrisankar.sreeram@cpt.univ-mrs.fr}}
\author[5]{Jacopo Taddei\thanks{jacopo.taddei@studio.unibo.it}}

\affil[1]{Aix-Marseille Univ, Université de Toulon, CNRS, CPT, Marseille, France}
\affil[2]{Physics Program, Bard College, 30 Campus Road, Annandale-on-Hudson, NY 12504, USA}
\affil[3]{The Rotman Institute, Western University, London ON, Canada}
\affil[4]{Perimeter Institute, 31 Caroline Street North, N2L 2Y5 Waterloo ON, Canada}
\affil[5]{Università di Bologna, Bologna, Italy}

\date{\today}

\date{\today}

\maketitle

\begin{abstract}\noindent
Tunneling processes offer a promising path for finding signatures of quantum gravity. While tunneling of geometry has long been recognized in the literature, few detailed analyses in covariant Loop Quantum Gravity have been carried out. We investigate spinfoam transitions in the holonomy representation, which naturally encodes the extrinsic curvature of boundary states. To reduce technical complications to a minimum, we study these amplitudes within the simple framework of the Ponzano–Regge spinfoam model for three-dimensional Euclidean quantum gravity. We identify the geometries dominating the spinfoam path integral in the classically forbidden regime when formulated in terms of dihedral angles as boundary data. We characterize these non-classical geometries and show that their contributions to the spinfoam amplitude are exponentially suppressed in the semiclassical limit via analytic continuation of the discrete gravity action. We argue that they satisfy all the desired properties of tunneling processes. We also shed light on quantum black-to-white-hole transitions, in particular clarifying the origin of the exponential suppression of various quantum amplitudes, while at the same time laying the basis for a future complete calculation of the amplitude in covariant Loop Quantum Gravity. 
\end{abstract}



\section{Introduction}
\label{sec:intro}

Tunneling is among the quintessentially quantum phenomena. Already playing a role in Hund's work on molecular spectra in 1926, it has gone on to significantly impact quantum technologies, from semiconductors to scanning tunneling microscopy \cite{Merzbacher:2002}, and foundational quantum questions, such as the time duration of tunneling events \cite{ramos2020measurement,sharoglazova2025energy}. As often formulated, one thinks of a particle or atom tunneling through a potential barrier.  However, this decomposition into a purely background structure, like a potential barrier, and a single degree of freedom of interest evolving on it, like a particle, is less obvious, and may not be possible in general, in the context of gravitational physics. Nevertheless, quantum gravitational tunneling can be made sense of and holds the potential to uncover peculiar quantum signatures of quantum gravity. 

\smallskip

One path to understanding quantum gravitational tunneling, which has a long history, is to study symmetry-reduced models. This has been pursued extensively in quantum cosmology, where Vilenkin's tunneling proposal \cite{Vilenkin1982} and the Hartle-Hawking no-boundary proposal \cite{HartleHawking1983} have provided models for cosmological origins without a big-bang singularity, for a recent review see \cite{Lehners2023}. For recent work on the no-boundary proposal see \cite{Feldbrugge:2017PRL,Feldbrugge:2017PRD}, and for a discussion of limitations and extensions, e.g., \cite{Feldbrugge:2018gin}. In this approach, there is a symmetry that allows one to introduce an effective potential and, using this potential, recover an analogy to the standard approach to tunneling. These works are in part rooted in Coleman, Callan, and De Luccia's investigation of quantum field theoretic tunneling \cite{Coleman1977,ColemanCallan1977}, including gravitational feedback \cite{ColemanDeLuccia}.  

\smallskip


A different and complementary viewpoint arises from discrete path integral approaches to quantum gravity, initiated by Ponzano and Regge \cite{Regge:1968} in three Euclidean dimensions and later extended to general spinfoam models. Rather than relying on a symmetry that isolates a single effective degree of freedom and an associated potential, these frameworks aim to implement the gravitational path integral directly as a sum over discrete geometries. In the original
work, Ponzano and Regge observed a remarkable correspondence between the Wigner ${6j}$ symbol, an SU(2)-invariant object, and discrete general relativity in three Euclidean dimensions.
Already in that first analysis, they highlighted that their construction naturally accommodates non-classical evolutions between distinct classical geometries. In particular, their semiclassical treatment showed that configurations with negative squared volume contribute exponentially suppressed quantum amplitudes, and yet their contributions do not vanishes. While they commented on the parallelism with WKB techniques and Feynman path integrals, the interpretation of their model as a genuine path-integral formulation of gravitational degrees of freedom quantized in the spirit of loop quantum gravity only emerged later \cite{Rovelli:1993kc}.
As a consequence, they did not identify these non-classical contributions explicitly as tunneling.
The connection between classically forbidden geometries and tunneling phenomena has been clarified more recently \cite{Dona:2024rdq}. In close analogy with quantum-mechanical tunneling, these works show that spinfoam amplitudes naturally receive contributions from classically forbidden configurations. Tunneling of quantum geometries is not an artifact of symmetry reduction or of introducing an external potential. Instead, it emerges intrinsically from the discrete gravitational path integral itself.

\smallskip

Modern spinfoam models build on these insights. Constructions such as the Barrett–Crane and Engle-Pereira-Rovelli-Livine (EPRL) models generalize the Ponzano–Regge idea to four dimensions and Lorentzian signature, providing a fully background-independent setting in which quantum gravitational tunneling can be studied. Here, spacetime geometry is fundamental, and a sum over histories of quantum geometry replaces the notion of degrees of freedom experiencing a potential barrier. The resulting amplitudes incorporate both classically allowed configurations and genuinely quantum contributions associated with classically forbidden ones. Despite this conceptual richness, tunneling in spinfoam models remains largely unexplored. Progress has been made in some interesting directions.  Lefschetz-thimble techniques \cite{Han:2020npv, Han:2021kll}, for instance, offer powerful methods for evaluating oscillatory integrals and capturing non-perturbative effects using sophisticated contour deformation and complex critical point techniques.  In parallel, several geometrically motivated approaches, e.g. \cite{DittrichPadua2024,Borissova:2024pfq,Borissova:2024txs}, have investigated the role of classically forbidden configurations in Lorentzian 4D geometries. Together, these developments further indicate the importance of studying tunneling. 

\smallskip

The present paper contributes to this direction. Our goal is to uncover the mechanism governing tunneling processes in spinfoam theories, thereby clarifying the semiclassical structure of loop quantum gravity and exploring its potential phenomenological consequences.
Because spinfoam models are mathematically demanding, meaningful progress requires three ingredients: complete control of the quantum amplitudes, a deep understanding of the underlying classical theory, and the ability to distinguish classically allowed from classically forbidden trajectories. For state-of-the-art four-dimensional models, many of these ingredients are still under development. This motivates our decision to work in the simpler setting of three-dimensional Euclidean quantum gravity, where the spinfoam model is fully understood and provides an ideal laboratory for defining and testing the relevant concepts.

\smallskip

While some of the technical elements we employ are present in the literature, the novelty of our approach lies in how we analyze them and in the geometric interpretation we extract.
A first step in this direction was taken in \cite{Dona:2024rdq}, which investigated tunneling in the length representation, with boundary states defined as eigenstates of length operators. Here we complement and extend that analysis by clarifying the tunneling mechanism in terms of the dual, holonomy- or angle-based, variables, a perspective that had not previously been explored.
This dual description provides new insights into how classically forbidden geometries appear in the quantum amplitude and how they contribute to the spinfoam path integral. The angle representation is not only of conceptual interest but also essential for certain physical applications.

\smallskip

In particular, the proposed black-to-white-hole transition in loop quantum gravity (a potential phenomenological signal of quantum gravity) requires a detailed understanding of how extrinsic curvature inverts across the transition quantum region. Since this inversion is naturally encoded in the conjugate, angle-like variables, a clear formulation of spinfoam tunneling in the dual representation is indispensable for a detailed study of this process.

\smallskip

Although our analysis is performed in three-dimensional Euclidean quantum gravity, the lessons we draw are broader. They can inform future work in the more physically relevant four-dimensional Lorentzian spinfoam models. Taking seriously the hypothesis that spinfoams recover discrete gravity in the semiclassical limit, with canonical LQG boundary states, we expect quantum evolution to be dominated by classically allowed geometries composed of spacelike boundary 4-simplices. Classically forbidden geometries, corresponding to Euclidean 4-simplices obtained via analytic continuation, should then play the role of tunneling configurations.  By understanding how these configurations emerge and contribute in three dimensions, we lay the groundwork for analyzing tunneling processes in full four-dimensional quantum gravity.


\smallskip


The paper is organized as follows: In Section~\ref{sec:trap}, we explain why tunneling in spinfoam can be used to compute the tunneling of a black hole to a white hole and why using coherent boundary states is not optimal for studying tunneling in quantum geometry. In Section~\ref{sec:PonzanoRegge}, we review the Ponzano-Regge spinfoam model and its description of spin networks in both spin and holonomy bases. In Section~\ref{sec:tetra}, we analyze classical transitions in three-dimensional Regge calculus, introducing the angle Gram matrix and discussing the classically allowed and forbidden evolution of surfaces. In Section~\ref{sec:amplitude}, we derive the semiclassical limit of the vertex amplitude in the holonomy basis, showing how the dynamics of classically allowed and forbidden transitions emerge naturally through the analytic continuation of the Regge action. We conclude with Section~\ref{sec:discussion}, where we discuss the implications of our results for modeling black-to-white hole transitions in covariant loop quantum gravity, and outline possible future directions.
Finally, an important note: To maintain the streamlined logic of the main text, we have moved all detailed calculations to the Appendices. These should not be viewed as mere review material. They form an integral part of the paper, providing essential derivations and supporting our arguments.

\section{The black-to-white hole transition and the problem with using coherent states}
\label{sec:trap}

A striking astrophysical discovery of the past few decades is the abundance and diversity of the objects that we call black holes. Direct and indirect observations of these objects are impeccably accounted for by classical general relativity. However, classical general relativity alone does not capture the physics of their extreme–curvature regions, where true quantum gravity effects are likely to become important. Studying these quantum regions is needed to understand what happens to a black hole at the end of its Hawking evaporation. 

An idea for describing the dynamics of these regions is to study the possibility of a tunneling of the black hole into a white hole (a time-reversed version of the black hole) \cite{Rovelli:2014cta, Rovelli:2017zoa, Haggard:2014rza, DeLorenzo:2015gtx, Christodoulou:2016vny, Bianchi:2018mml, Ashtekar2018, Lewandowski:2022zce, Giesel:2022rxi, Martin-Dussaud:2025qtr}. This scenario offers a natural solution to the black hole information ``paradox'' by erasing it entirely, is empirically testable at least in principle \cite{Christodoulou:2023hyt}, and provides a natural candidate for dark matter without the need for any new physics (such as new fields, particles, or modifications of the field equations) \cite{Carr:2020xqk, Green:2020jor}. 

\smallskip

Covariant Loop Quantum Gravity offers a background-independent, Lorentzian path-integral framework that is ideally suited to describe such spacetime quantum transitions \cite{Rovelli2004}. Numerous efforts have been made to develop a concrete implementation of the black-to-white-hole transition amplitude in Covariant LQG, first using substantial simplifications and analytical methods \cite{Christodoulou:2018ryl,DAmbrosio:2020mut, Soltani:2021zmv, Christodoulou:2023psv} and then confirming the results with numerical calculations \cite{Frisoni:2023agk,Han:2024rqb}. 



\smallskip

Generally, tunneling appears in the semiclassical limit of the amplitudes of transitions between boundary states that are classically forbidden and can be understood mathematically in terms of complex saddle points in a sum over histories analytically continued in the complex plane.  In gravity, a simple realization of this phenomenon was studied in the Ponzano-Regge model, a Euclidean 3D spinfoam theory \cite{Dona:2024rdq}: analytically continued classical geometries govern these amplitudes, each contributing an exponential suppression factor given by the Regge action, the discrete counterpart of Einstein's action for general relativity \cite{Asante:2021phx,Dittrich:2023rcr,Bojowald:2021cqg,Motaharfar:2022pjp}.

\smallskip

Early work in full, Lorentzian loop quantum gravity has observed a similar exponential suppression of the amplitude for boundary states defined by coherent states peaked on the spacelike boundary geometry, and attributed this suppression to a tunneling process. However, as we point out below, relying on coherent boundary states to study tunneling amplitudes can lead to misinterpretation. Non-classical geometries dominate tunneling amplitudes, whereas coherent states are peaked on classical geometries. Here, we argue for a reinterpretation of the suppression and consider improvements to those computations.

\smallskip

The black-to-white hole transition is characterized by a flip in the sign of the extrinsic curvature between the two spacelike boundary hypersurfaces of the black and white hole regions, respectively \cite{Haggard:2014rza}. We notice that describing tunneling transition amplitudes purely in terms of {\em intrinsic} geometry obscures this key aspect of the physics. It is more convenient to move to the conjugate (holonomy) basis, and this is what we do here. 

A helpful analogy to understand this fact is the quantum reflection of a particle from a potential barrier, when the energy of the particle {\em exceeds} the height of the barrier \cite{Jaffe:2010}. Classically, a particle gets past the barrier, while quantum mechanically, there is a non-vanishing probability of reflection. In the position representation, the description of this phenomenon is cumbersome and requires complex coordinates.
In contrast, the momentum representation displays the reflection naturally as a reversal in the sign of the momentum. Complex positions emerge naturally as the conjugate variables associated with classically forbidden paths described in terms of real momenta. Likewise, transitions between geometries distinguished by their extrinsic curvature become clearer in the holonomy basis, where they are directly encoded. 

The use of extrinsic coherent states in Loop Quantum Gravity (coherent spin networks) is advantageous when constructing kinematic states that approximate both the intrinsic and extrinsic geometry of a spacelike hypersurface \cite{Thiemann:2002vj, Bianchi:2009ky}. However, in studying the tunneling of quantum geometries with truncated degrees of freedom and a minimal two-complex, this strategy turns out to be suboptimal. The geometries dominating the transition amplitude are non-classical, and the calculation of coherent states becomes excessively complicated. 
This parallels one-dimensional quantum mechanics, where many analytical results on reflection and transmission coefficients for plane waves exist and are transposed to wave packets, which are linear superpositions of plane waves. In the spinfoam framework, however, we lack the analogue of the plane-wave calculation and have directly considered coherent states (wave packets). A genuine holonomy eigenstate basis analysis provides a clearer picture and should be preferred over the coherent-state treatment.
Some current literature on quantum geometry tunneling \cite{Christodoulou:2018ryl,DAmbrosio:2020mut, Soltani:2021zmv, Christodoulou:2023psv} employs coherent states, unintentionally concealing the physical mechanism in technical details, which also complicates interpretation.

\smallskip

These calculations of the amplitude of the black-to-white hole transition were performed in a minimal discretization. To simplify the calculations, they used the simplest triangulation consistent with a classically allowed, albeit degenerate, evolution. The calculation led to an exponential suppression of the amplitude. At first, this seems reasonable. However, this is misleading: the suppression arises from an artificial mismatch between the extrinsic geometry encoded in the boundary coherent state and the extrinsic geometry that is compatible with the intrinsic boundary data. Let us illustrate this fact with a simple example.

\smallskip

Consider a quantum particle whose dynamic is given by an action $S$. The quantum transition between initial and final states, described by the positions $x_0$ and $x_1$, is given by the propagator $K(x_0,x_1)$, which can be expressed in terms of Feynman's path integral. In the semiclassical limit, the transition amplitude is dominated by classical paths. These are solutions of the equations of motion of the underlying classical theory, and the transition amplitude between the initial and final states ($x_0$ and $x_1$) reduces to \cite{Morette:1951zz,DeWitt-Morette:1976ydh,Carlitz:1984ab}
\begin{equation}
\label{eq:pathintegral}
\mathcal{K}(x_0,x_1) = \int_{x_0}^{x_1} \mathcal{D}[q] e^{ \frac{i}{\hbar}S(q)} \approx \sqrt{\frac{i}{2\pi \hbar}\frac{\partial^2 S_c}{\partial x_0\partial x_1}} e^{\frac{i}{\hbar}S_c(x_0,x_1)} \ ,
\end{equation}
where $S_c(x_0,x_1)$ is Hamilton's principal function, i.e., the action evaluated on a solution of the equations of motion compatible with the boundary conditions. For simplicity, we assume there is only one such classical trajectory. 

If instead of computing the transition between two eigenstates of position $x_0$ and $x_1$, we want to compute the amplitude of the transition between two coherent states, we have to consider the boundary states peaked at the classical positions $q_0$ and $q_1$ and momenta $p_0$ and $p_1$,

\begin{equation}
\label{eq:coherent-state}
    \psi_{q_0, p_0,q_1,p_1}(x_0,x_1)=\left(\frac{1}{\hbar\pi}\right)^{\frac12} e^{-\frac{1}{2\hbar}(x_0-q_0)^2}e^{\frac{i}{\hbar}p_0(x_0-q_0)}e^{-\frac{1}{2\hbar}(x_1-q_1)^2}e^{-\frac{i}{\hbar}p_1(x_1-q_1)} \ .
\end{equation}
Contraction of the propagator \eqref{eq:pathintegral} and the boundary coherent state \eqref{eq:coherent-state} gives the amplitude 

\begin{equation}
\label{eq:coherent-amplitude}
    A(q_0, p_0,q_1,p_1) = \int \int \d x_0\, \d x_1\, \mathcal{K}(x_0,x_1) \, \psi_{q_0, p_0,q_1,p_1}(x_0,x_1) \ .
\end{equation}

This amplitude can be simplified in the semiclassical limit: First,  approximate the propagator with its semiclassical form. Then, perform the integral over the boundary Hilbert space using the saddle point approximation. The result is
\begin{equation}
    A(q_0, p_0,q_1,p_1) \approx \left(\frac{1}{\hbar\pi}\right)^{\frac12}\sqrt{\frac{2\pi\hbar \,i}{H}\frac{\partial^2 S_c}{\partial q_0\partial q_1}} \  e^{\frac{i}{\hbar}S_c(q_0,q_1)}
     e^{-\frac{1}{2 \hbar}(p_0+\partial S_c/\partial q_0)^2} e^{-\frac{1}{2 \hbar}(p_1-\partial S_c/\partial q_1)^2} \ ,
    \label{eq:generalpathintegral}
\end{equation}
where $H = \left(\frac{\partial^2 S_c}{\partial q_0\partial q_1}\right)^2+\left(1-i\frac{\partial^2 S_c}{\partial q_0^2}\right)\left(1-i\frac{\partial^2 S_c}{\partial q_1^2}\right)$ is the Hessian of the integration over the boundary Hilbert space. The amplitude oscillates with $S_c(q_0,q_1)$, Hamilton's principal function evaluated on the classical trajectory joining the classical labels of the boundary state $q_0$ to $q_1$. Moreover, the amplitude is exponentially suppressed unless the boundary momenta $p_0$ and $p_1$ agree with the classical momenta computed from Hamilton's principal function, $p_0 = -\partial S_c/\partial q_0 $ and $p_1 = \partial S_c/\partial q_1$. This suppression is quadratic in the momenta.  

\smallskip

In a spinfoam theory, the result is completely analogous. We carry out the calculation using the Ponzano-Regge model for simplicity, but one should obtain the same result also in the Lorentzian EPRL model \cite{Bianchi:2009ky}. The main differences lie in how the intrinsic geometry is encoded using the intertwiner degrees of freedom and the areas of the triangles. With coherent boundary states peaked on the lengths $\ell_f$ and dihedral angles $\xi_f$, the transition amplitude is peaked on a classical geometry (we ignore the orientation ambiguity and all the Hessians for simplicity)
\begin{equation}
    A(\ell_f, \xi_f) \approx e^{\frac{i}{\hbar}S_R(\ell_f)}
     e^{\sum_f-\frac{1}{2 \hbar}(\xi_f-\partial S_R/\partial \ell_f)^2} \ ,
    \label{eq:spinfoam}
\end{equation}
where $S_R = \sum_f \ell_f \psi_f(\ell_f)$ is the Regge action (Hamilton's principal function of Regge gravity) and $\psi_f=\partial S_R/\partial \ell_f$ are the dihedral angles. If the extrinsic curvature compatible with flat three-dimensional classical dynamics is (partially) degenerate 
($\psi_f=0$), but the corresponding boundary momenta $\xi_f$ do not vanish, the amplitude is suppressed by $e^{-\xi_f^2/(2\hbar)}$. This suppression is unrelated to tunneling. By contrast, in genuine tunneling of quantum geometries \cite{Dona:2024rdq} one obtains a suppression $e^{-\frac{1}{\hbar} \ell_f |\psi_f(\ell_f)|}$ after analytically continuing the dihedral angles to complex values. This contribution comes from the dynamically primary Regge action.

\section{The Ponzano-Regge spinfoam model}
\label{sec:PonzanoRegge}
The Ponzano-Regge spinfoam model provides a path-integral quantization of three–dimensional Euclidean gravity. It is defined on a simplicial 2-complex, where one assigns transition amplitudes to three–valent spin network states living on the boundary of the complex. These boundary states describe two-dimensional quantum surfaces tiled by Euclidean triangles, with each triangle dual to a (three-valent) node of the spin network. The overall spinfoam amplitude encodes the quantum evolution of these surfaces and factorizes into a product over vertex amplitudes, each of which is associated with a vertex of the 2-complex (these, in turn, are dual to tetrahedra). In this section, we recall only the minimal ingredients needed. For a more detailed exposition, see the original paper \cite{Regge:1968} and the extensive literature that followed \cite{Roberts:1998zka,Freidel:2004vi,Freidel:2005bb,Barrett:2008wh,Rovelli:2014ssa}.

\smallskip

\begin{figure}
    \centering
    \includegraphics[width=0.4\linewidth]{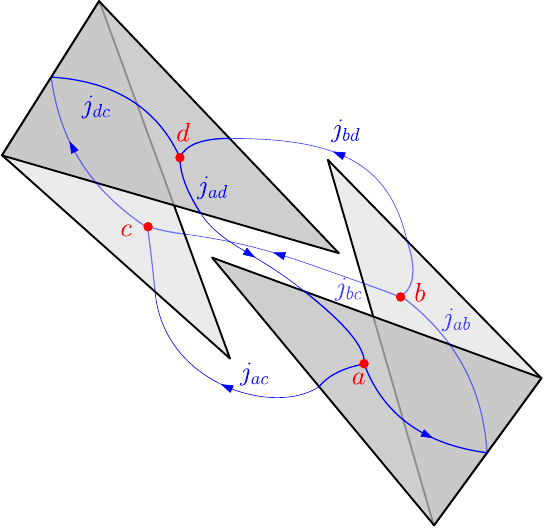}
    \includegraphics[width=0.4\linewidth]{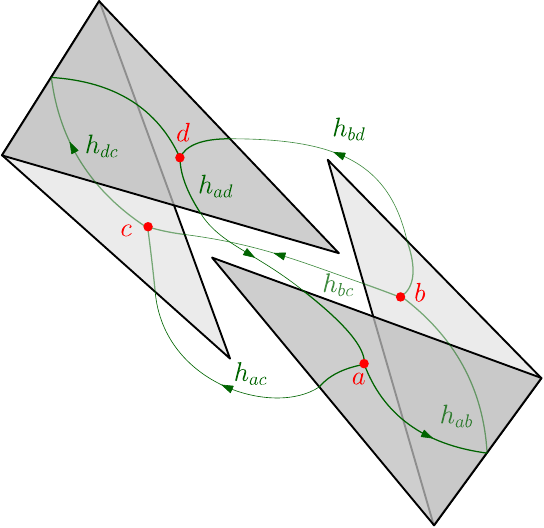}
    \caption{Canonical evolution of a surface discretized by two triangles, shown at an initial stage (left pair) and a final stage (right pair), that is, you can think of ``time'' as flowing from left to right in each panel. \textit{Left panel} – the boundary spin network is highlighted in blue, with spin labels $j_{ab}$. \textit{Right panel} – the boundary spin network is highlighted in green, with holonomy labels $h_{ab}$. 
\label{fig:spinnetwork}}
\end{figure}

At each spinfoam vertex, the boundary spin network forms a tetrahedral graph as shown in Figure~\ref{fig:spinnetwork}. We label its four nodes (each dual to a boundary triangle) by $a=1,\dots,4$, and each oriented link by the ordered pair $ab$ carrying spin $j_{ab}$. To each link we also associate an $SU(2)$ holonomy $h_{ab}$. The boundary state is then given by 
\begin{equation}
    \label{eq:spin_network}
    \Psi_{v}^{j_{ab}}(h_{ab})=\bigotimes_{n} i_{n} \bigotimes_{(ab)} D^{j_{ab}}(h_{ab}) \ ,
\end{equation}
where we have omitted the magnetic indices that are contracted and summed over using the intertwiners $i_n$, and following the connectivity of the graph in Figure~\ref{fig:spinnetwork}.
The spins $j_{ab}$ determine the lengths $\ell_{ab} \;=\; \ell_P\,\sqrt{j_{ab}(j_{ab}+1)}$ of the edge shared by the boundary triangles $a$ and $b$ in terms of the Plank length $\ell_P \;=\;\hbar G$. The holonomies $h_{ab}$ encode the extrinsic curvature of the surface via parallel transport from triangle $a$ to triangle $b$. In what follows, we set the Planck length to unity $\ell_P=1$.

\smallskip

A convenient parametrization of the holonomies is provided by the \emph{twisted–geometry} framework. In the twisted–geometry description of Loop Quantum Gravity holonomies \cite{Freidel:2010aq,Freidel:2010bw,Dupuis:2012yw}, one introduces complex spinors $\ket{z_{ab}}$ attached to each oriented link $(ab)$ and an angle $\phi_{ab}$ encoding the extrinsic dihedral angle. We restrict ourselves, without loss of generality, to unit spinors to simplify the notation. One then writes
\begin{equation}
  \label{eq:param}
  h_{ab}
  \;=\;
  e^{i\frac{\phi_{ab}}{2}}\,\sket{z_{ba}}\bra{z_{ab}}
  \;-\;
  e^{-i\frac{\phi_{ab}}{2}}\,
     \ket{z_{ba}}\sbra{z_{ab}}\ .
\end{equation}
Here, the spinors $\ket{z_{ab}}$ carry the intrinsic geometry (in 3D, they encode the shape of the triangles). The angles $\phi_{ab}$ carry the extrinsic geometry (the dihedral angles between adjacent triangles). The spinorial notation is explained in detail in Appendix~\ref{app:semiclassicalspinnetwork}.

\smallskip

The Ponzano-Regge vertex amplitude in the spin network basis is given by the Wigner $6j$-symbol \cite{Varshalovich:1988krb}
\begin{equation}
    \label{eq:Avspins}
    A_v (j_{ab})=\SixJ{j_{12}&j_{13}&j_{14}}{j_{34}&j_{24}&j_{23}} \ .
\end{equation}
Spins and holonomies $\bigl(j_{ab},h_{ab}\bigr)$ are conjugate variables on the Hilbert space of LQG. The vertex amplitude in the holonomy basis is related to \eqref{eq:Avspins} by a (non-commutative) Fourier transform on the LQG Hilbert space using the spin network basis \eqref{eq:spin_network}
\begin{equation}
    \label{eq:Avhol}
    A_v(h_{ab})
    \;=\;
    \sum_{j_{ab}} A_v(j_{ab}) \Psi_{v}^{j_{ab}}(h_{ab}) \ .
\end{equation}

The sum over spins can be computed exactly and gives the locally flat connection condition that imposes trivial parallel transport within the vertex \cite{Livine:2007mr,Dona:2022hgr}. Technically, this is implemented with delta functions, which is very convenient in some circumstances, but obscures how geometry emerges from the asymptotics. So, we will proceed using the formulation of \eqref{eq:Avhol}. 

This completes our brief review of the quantum model. We turn now to classically allowed and forbidden transitions.

\section{Classically allowed transitions}
\label{sec:tetra}
In the path integral formalism, transition amplitudes in the semiclassical regime are dominated by classical paths, as in \eqref{eq:pathintegral}. The classical theory underlying the Ponzano-Regge spinfoam model is three-dimensional Euclidean Regge gravity. Therefore, before analyzing the semiclassical regime of the spinfoam model, we will briefly review three-dimensional Euclidean Regge gravity. In the first order formulation of Regge calculus \cite{Barrett:1994nn}, the canonical variables are the edge-lengths $\ell_{ab}$ of the triangulation and the 3D dihedral angles $\phi_{ab}$ dual to these edges. They are canonically conjugate to one another.

\smallskip

The equations of motion fix the dihedral angles as functions of the six edge lengths of a Euclidean tetrahedron $\phi_{ab} = \psi_{ab}(\ell_{ab})$. We will use the symbol $\phi_{ab}$ to denote the dihedral angles treated as independent variables from the lengths, and $\psi_{ab}(\ell_{ab})$ to represent the function of the lengths that describes the dihedral angles of a Euclidean tetrahedron \cite{Dona:2024rdq}. Together with the equations of motion, these functions also ensure that the sum of the dihedral angles around a bulk edge is $2\pi$, which is consistent with the flatness of 3D gravity. Regge calculus, when interpreted as a canonical theory, describes the evolution of two-dimensional surfaces through Hamilton's principal function \cite{Rovelli:2014ssa, Dittrich:2011ke,Bahr:2009qd}. Each surface consists of a collection of triangles joined along their edges. The classical dynamics of three-dimensional canonical gravity can then be understood as a sequence of local moves, where \textit{Euclidean tetrahedra} are glued onto the surface in all possible ways \cite{Dittrich:2011ke}, resulting in a multi-fingered evolution.

\smallskip

Hamilton's Principal function is the classical Regge action evaluated on the classical trajectory of a Euclidean tetrahedron with edge lengths $\ell_{ab}$ and dihedral angles $\psi_{ab}$\footnote{We use external dihedral angles to align with the quantum gravity literature. However, the angles conjugate to the lengths are the internal ones. Nothing changes in practice, but this distinction should be kept in mind.}
\begin{equation}
S_R(l_{ab}) = \sum_{ab} \psi_{ab}(\ell_{ab}) \ell_{ab}\ .
\label{eq:PHF}
\end{equation}

This description prefers lengths over angles. Can we reverse the perspective and use dihedral angles as the preferred variables instead? Dihedral angles of a Euclidean tetrahedron are not independent: they satisfy a constraint that is local to each tetrahedron, and this constraint must be incorporated into the action. For each tetrahedron, we define the angle Gram matrix as
\begin{equation}
\label{eq:Gram}
G_{ij}(\phi_{ab}) = \cos{\phi_{ij}}, \quad i,j=1,\dots,4\ ,
\end{equation}
with the convention $\phi_{ii} = 0$.
The Gram matrix of a Euclidean tetrahedron is singular—its rank is 3, and, hence, its determinant vanishes. The null space of the Gram matrix is spanned by a positive vector proportional to the area vector, whose components are the areas of the faces of the tetrahedron. This is a consequence of the closure condition. (See Appendix~\ref{app:Gram} for a simple derivation and further characterization.) We impose the closure constraint 
\begin{equation}
\label{eq:closure}
\det (G(\phi_{ab})) = 0
\end{equation}
via a Lagrange multiplier \cite{Barrett:1994nn}. Focusing on a single tetrahedron (each tetrahedron has an independent constraint), the first-order action reads:
\begin{equation}
S_R(\ell_{ab},\phi_{ab}) = \sum_{ab} \ell_{ab} \phi_{ab} + \tilde\lambda \det (G(\phi_{ab})) \ .
\label{eq:FOAction}
\end{equation}
The equation of motion for the dimensionful Lagrange multiplier $\tilde\lambda$ enforces the closure constraint \eqref{eq:closure}. The equations of motion for the angles give an expression for the edge lengths as functions of the angles \cite{Barrett:1994nn, Asante2018degrees}\footnote{We started following the derivation of the derivative of \cite{Asante2018degrees}. Still, we completed it in a different way to get a fully geometrical proportionality factor between the Gram matrix cofactor and $A_aA_b$. Our derivation is in Appendix~\ref{app:Gram}.}
\begin{align}
\frac{\delta S_R}{\delta \phi_{ab}} = \ell_{ab} - \tilde\lambda \frac{81}{4}\frac{V^4}{(A_1A_2A_3A_4)^2} A_a A_b \sin \phi_{ab}=
\ell_{ab} - \lambda \, l_{ab}( \phi_{ab})  \ .
\label{eq:eomangles}
\end{align}
In the second equality, we have used the formula
\begin{equation}
l_{ab}(\phi_{ab}) = \frac{2}{3} \frac{A_a A_b}{V} \sin \phi_{ab} 
\label{eq:lengthform}
\end{equation}
for the lengths in terms of the volume of the tetrahedron $V$ and the areas $A_b$, and introduced the dimensionless scale 
\begin{equation}
    \lambda \equiv \tilde\lambda \, \frac{3^5}{2^3}\frac{V^5}{(A_1A_2A_3A_4)^2} \ .
\end{equation} 
We will denote as $l_{ab}(\phi_{ab})$ the geometric length of the tetrahedron thought of as a function of the dihedral angles. Note that the lengths cannot be uniquely determined as an arbitrary overall scale remains. This is expected: rescaling the tetrahedron multiplies all edge lengths by the same factor, while leaving the dihedral angles invariant. Hence, the dihedral angles determine the shape of the tetrahedron up to a global dimensionless scale $\lambda$. 

\smallskip

The action \eqref{eq:FOAction} reduces to Hamilton's principal function in terms of the dihedral angles when evaluated on-shell, i.e., on solutions to the equations of motion
\begin{equation}
S_R(\phi_{ab}) = \lambda \sum_{ab} l_{ab}(\phi_{ab}) \phi_{ab}  \ .
\label{eq:FOActionPhi}
\end{equation}
We slightly abuse notation by employing the same symbol for both \eqref{eq:FOActionPhi} and \eqref{eq:PHF}. Their arguments distinguish them. Note that the two Hamilton's principal functions are related, as they describe the same evolution and, thus, the same canonical transformation. Equation \eqref{eq:FOActionPhi} can also be derived from \eqref{eq:PHF} by performing a Legendre transformation
\begin{equation}
S_R(\phi_{ab}) =   \sum_{ab} \ell_{ab} \phi_{ab} + \sum_{ab} \ell_{ab}(\phi_{ab}) \psi_{ab} \ ,
\end{equation}
where we formally invert $\ell_{ab}$ in terms of $\phi_{ab}$ using the equation
\begin{equation}
\label{eq:eomref}
\phi_{ab} = \psi_{ab}(\ell_{ab}) \ , \quad \text{which is equivalent to}  \quad \ell_{ab} = \lambda \, l_{ab}(\phi_{ab}) \ ,
\end{equation}
with an arbitrary dimensionless scale factor $\lambda$.

\smallskip
The constraint \eqref{eq:closure} is a bit subtle. Consider a triangle described by its edge lengths. If the prescribed edge lengths satisfy the triangle inequalities, this guarantees the existence of a boundary triangle with positive area. However, satisfying the triangle inequalities for all four triangles is not sufficient to ensure that they can be assembled into a Euclidean tetrahedron. Similarly, in the dual picture, a set of dihedral angles satisfying the closure constraint \eqref{eq:closure} and whose associated Gram matrix has a positive area vector constitutes a necessary—but not sufficient—condition for the existence of a Euclidean tetrahedron with those angles.

We recall that triangles that can form a Euclidean tetrahedron correspond to a classically allowed evolution. If they do not, but still satisfy the closure constraint \eqref{eq:closure}, they represent classically forbidden configurations. The squared volume is a useful quantity to distinguish between the two cases and has proven to be effective in the length picture \cite{Dona:2024rdq}. However, in the angle picture, it is not particularly convenient since it is not scale invariant.

How, then, can we discriminate between allowed and forbidden classical evolution? Given a Gram matrix $G$, the corresponding tetrahedron is embeddable in a metric space if $G$ is similar to the metric of that space (in Appendix~\ref{app:Gram}, we report a detailed proof inspired by \cite{Asante:2025qbr}). For a tetrahedron to be embeddable in Euclidean space, the corresponding Gram matrix constructed from the dihedral angles must be positive semi-definite with signature $(3,0,1)$.\footnote{We use the notation $(n_+,n_-,n_0)$, where $n_+$ is the number of positive eigenvalues, $n_-$ the number of negative ones, and $n_0$ the number of zero eigenvalues.} However, if a Gram matrix has signature $(2,1,1)$, then no Euclidean tetrahedron is compatible with that set of angles, although a Lorentzian one exists. The signature of a Gram matrix, defined by the dihedral angles associated with a boundary surface and having one positive null vector, fully determines whether there is a compatible classical evolution.

\smallskip

To illustrate this in detail, we proceed with a concrete example. To simplify the analysis, we consider a symmetric geometry. The results are general, but for the sake of concreteness, we reduce the study to a two-parameter family of transitions. Specifically, we examine the evolution between two surfaces, each discretized by two triangles with non-trivial extrinsic curvature represented by the dihedral angle between the triangles. We prescribe the angles so that the areas of the triangles are equal in pairs. For simplicity, we take these angles to be equal, setting $\phi_{12} = \phi_{34}$. Between the two surfaces, there are four additional dihedral angles. We impose a symmetry by setting one pair of opposite dihedral angles equal ($\phi_{13} = \phi_{24}$), fix another ($\phi_{14} = \phi_R = \arccos\left(-\tfrac{1}{3}\right)$), and determine the last one ($\phi_{23}$) by requiring that the closure constraint \eqref{eq:closure} is satisfied. The closure condition is nonlinear and generally yields two solutions for $\cos(\phi_{23})$, corresponding to different angle configurations. For the purpose of the plot, we select the solution compatible with the presence of the regular tetrahedron where all the angles $\phi_{ab}=\phi_R$.

\begin{figure}[!ht]
    \centering
    \includegraphics[width=0.7\textwidth]{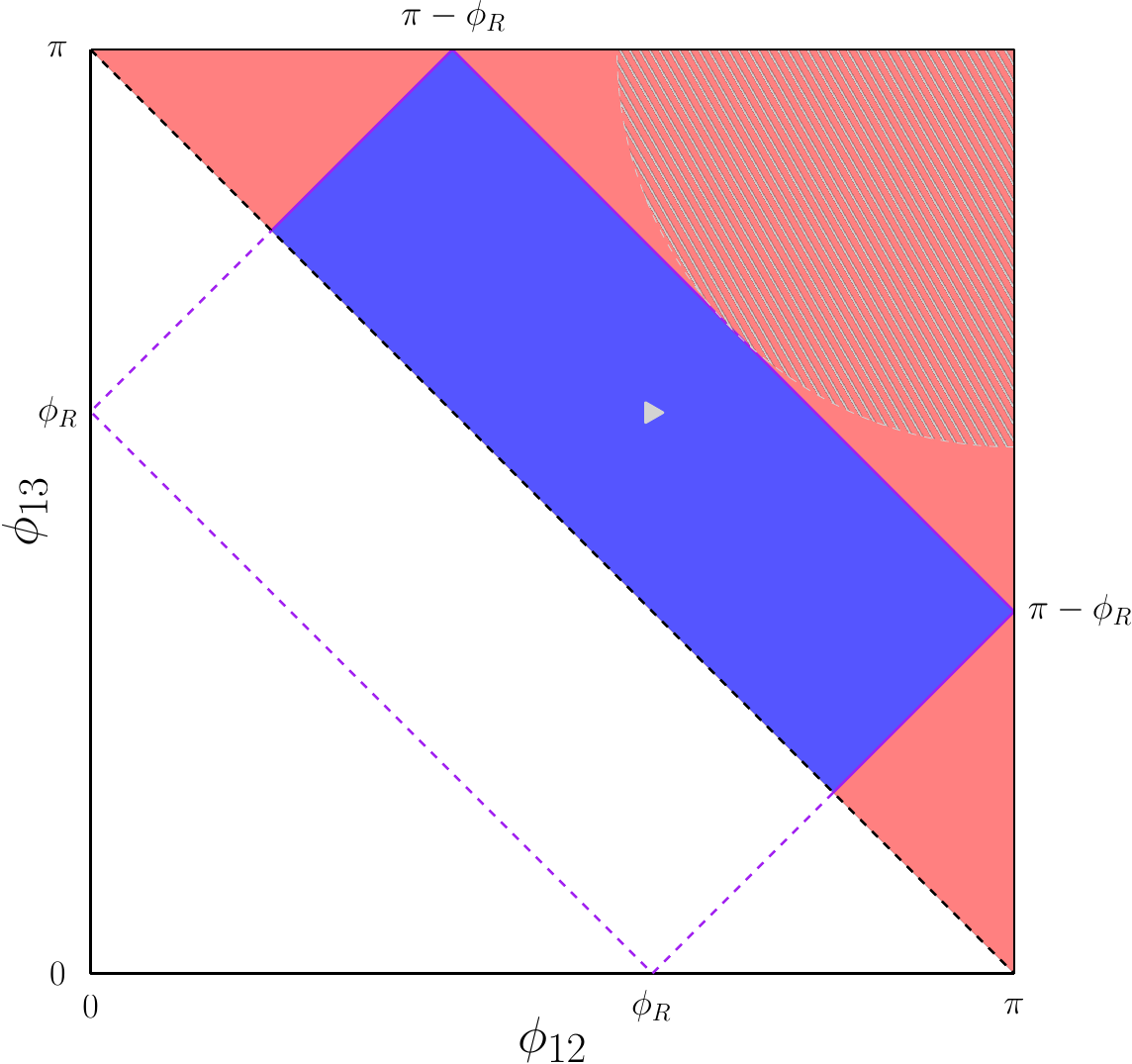}
    \caption{
        Classically allowed (blue) and forbidden transitions (red) between symmetric boundary data. We identify $\phi_{12} = \phi_{34}$ and $\phi_{13} = \phi_{24}$, fix $\phi_{14} = \phi_R= \arccos\left(-\tfrac{1}{3}\right)$, and determine $\phi_{23}$ using the closure condition. We select the solution compatible with the presence of the regular tetrahedron, highlighted by a gray triangle in the plot.
        The blue region corresponds to configurations where the transition is allowed (Euclidean signature), while the red background indicates classically forbidden configurations (Lorentzian signature). In the top-right corner, we shade the region where not all edge lengths are time-like and $l_{23}$ becomes spacelike. We extend the boundary between the two signature regions to include non-positive area vectors, shown as a dashed purple line.
    }
    \label{fig:transition-region}
\end{figure}

We then explore whether a classical transition is allowed for different values of the two free dihedral angles in this symmetric configuration: $\phi_{12}$ (identified with $\phi_{34}$) and $\phi_{13}$ (identified with $\phi_{24}$), both ranging from $0$ to $\pi$. We find that there exist entire regions in this configuration space of boundary surfaces where the transition is classically forbidden. See the plot in Figure~\ref{fig:transition-region}. The bottom-left part of the diagram is not allowed, as the Gram matrix constructed from these angles does not have a positive null eigenvector. In the classically allowed region, the Gram matrix has Euclidean signature, while in the classically forbidden region, it has Lorentzian signature.  At the boundary between the two regions, the Gram matrix has rank two, corresponding to a degenerate geometry. In the Lorentzian region, when all the angles are real, all the edge lengths are time-like (i.e., imaginary). In the top-right corner of the plot, we highlight a special Lorentzian region where the edge length $\ell_{23}$ becomes spacelike: the length is real while the corresponding angle $\phi_{23}$ is complex (getting the correct sign for $\ell_{23}$ requires some care in selecting the correct branch of $\phi_{23}$ as you transition between the shaded and unshaded forbidden regions). Additionally, along the anti-diagonal, two components of the area vector diverge. For readers interested in exploring similar setups, we provide a companion Mathematica notebook \cite{NotebookRepo} to experiment with alternative configurations. 

\smallskip

Expressing all of the geometrical quantities of a tetrahedron in terms of its dihedral angles is intricate. In particular, an explicit formula for any given geometrical quantity can only be defined up to an overall scale, which must be fixed in some way, and there is no obvious choice to make. In Appendix~\ref{app:Gram} we detail the space of dihedral angles giving rise to tetrahedra and explore two concrete ways of fixing this scale. Consistent with the examples above, both of these methods give lengths $l_{ij}$ that are imaginary whenever an edge is timlike, as expected for a Lorentzian tetrahedron.

\section{Semiclassical limit of the vertex amplitude}
\label{sec:amplitude}
In quantum mechanics, the semiclassical limit corresponds to a regime dominated by paths whose action is much larger than Planck's constant, $S \gg \hbar$. This holds in both the position and momentum pictures. Traditionally, we focus on the position one because the physical intuition is stronger, but the same analysis can be carried out in the momentum picture as well. A similar change of picture is also possible in the context of spinfoam models, as we show explicitly in this section.  This involves a shift in perspective: from viewing the transition amplitude as a function of the spins $j_{ab}$ (or lengths $\ell_{ab}$) to taking it to be a funciton of the holonomies $h_{ab}$ (or dihedral angles $\phi_{ab}$). 

\smallskip

 The semiclassical regime of spinfoam models corresponds to $S_R \gg \ell_P$, where $S_R$ is the Regge action and $\ell_P$ is the Planck length. Since the dihedral angles are dimensionless and bounded (in the Euclidean case), and we work in units where $\ell_p = 1$, this condition is typically associated with large spin values $j$. However, it is essential not to confuse this with merely large-scale geometries. Exploring the semiclassical regime in spinfoams does not necessarily imply studying large geometrical structures—it indicates that quantum fluctuations are suppressed relative to the classical action \cite{Carlitz:1984ab}. 

\smallskip

States in the three-valent LQG Hilbert space correspond to triangulated quantum surfaces. The quantum transition amplitude given in equation~\eqref{eq:Avhol} fully characterizes the quantum evolution of the boundary state in the holonomy basis. In the simplest case of evolution, the boundary state 2-complex is described by a trivalent tetrahedral spin network graph, as illustrated in Figure~\ref{fig:spinnetwork}. We compute the asymptotics of the transition amplitude \eqref{eq:Avhol} in several steps.

\smallskip

The semiclassical limit of the $6j$ symbol \eqref{eq:Avspins} has been extensively studied in the literature \cite{Regge:1968, Roberts:1998zka, Dupuis:2009sz, AquilantiEtAl2012, Dona:2019jab}. In the spin basis, the Ponzano-Regge vertex amplitude in the semiclassical regime is dominated by the geometry of a tetrahedron whose edge lengths are compatible with the boundary spins. The amplitude is well approximated by
\begin{equation}
{A}_v(j_{ab}) \approx \frac{1}{2\sqrt{-12 i \pi V(\ell_{ab})}} e^{i S_R(\ell_{ab})} + \text{inverted orientation} \ ,
\label{eq:semiclasstrans}
\end{equation}
where $S_R$ is interpreted as Hamilton's principal function defined in \eqref{eq:PHF}, i.e., the Regge action of the Euclidean tetrahedron described by the edge lengths $\ell_{ab}$. These lengths, in turn, are directly functions of the spins.

\smallskip

Next, we approximate the spin network states $\Psi_{v}^{j_{ab}}(h_{ab})$ of Equation~\eqref{eq:spin_network} in the semiclassical regime. We fix the holonomies to represent the parallel transport around the faces of a tetrahedron and use the twisted geometry parametrization given in \eqref{eq:param}. Each sum over magnetic indices in the contraction between Wigner matrices and three-valent intertwiners in \eqref{eq:spin_network} can be replaced by an integral over $SU(2)$ spinors, which form an overcomplete basis of the spin representation. We then approximate the integral using saddle point techniques, as detailed in Appendix~\ref{app:semiclassicalspinnetwork}. In summary, the integrals are peaked when the spinors are aligned with the holonomy spinors. The intertwiners contract with these spinors to form three-valent Livine–Speziale coherent intertwiners $c_n(z_{ab})$ (derived in \cite{Livine:2007vk}). These contributions are exponentially suppressed unless the closure condition is satisfied, which, in this simple case, requires that the spinors describe the shape of a triangle which is compatible with the one determined by the lengths $\ell_{ab}$ at that node. Otherwise, they scale as a power law in the spins and do not exhibit any oscillatory phase. Furthermore, the dihedral angles $\phi_{ab}$ are extracted from the holonomies and are encoded in the corresponding phases.
\begin{equation}
\label{eq:large_spin_network}
\Psi_{v}^{j_{ab}}(h_{ab}) \approx \mathcal{N}(\ell_{ab})e^{-i \sum_{ab}\ell_{ab} \phi_{ab}} + \text{inverted orientation} \ .
\end{equation}
The overall multiplicative factor $\mathcal{N}(\ell_{ab}) =   (2 \pi)^{12} \frac{\prod_n c_n(z_{ab})}{\prod_{ab} (\ell_{ab})^2}$ arises from the contribution of coherent intertwiners and the Hessians of the spinor integrals. The spinors $z_{ab}$ are determined by the closure conditions at the nodes and are functions of the length variables. The sum is over the spins, which ultimately are the fundamental variables, but in the semiclassical regime are simply related to the lengths ($\ell_{ab} = \sqrt{j_{ab}(j_{ab}+1)} \approx j_{ab} + \frac{1}{2}$).

\smallskip

Combining the asymptotics of the vertex amplitude \eqref{eq:semiclasstrans} and the spin network state \eqref{eq:large_spin_network} we obtain 
\begin{equation}
    A_v(h_{ab})
    \;\approx\;
     \sum_{j_{ab}}\frac{\mathcal{N}(\ell_{ab})}{2\sqrt{-12 i \pi V(\ell_{ab})}} \exp\left({i \sum_{ab}\ell_{ab}\psi_{ab}(\ell_{ab})} - i \sum_{ab} \ell_{ab} \phi_{ab}\right) + \text{all possible orientations}\ .
\end{equation}

\smallskip

To perform the final summation, we employ the Laplace method for discrete sums, also known as the saddle point approximation. It is typically used for integrals, but can also be employed for summations \cite{deBruijn:1981,BenderOrszag:1999}. A general overview of this technique is provided in Appendix~\ref{app:Laplace}. While this approach is mathematically equivalent to the Poisson summation formula often used in the spinfoam literature \cite{Engle:2020ffj,Han:2020npv,Han:2021kll}, it offers a more direct and intuitive route in our context. 
This perspective was developed initially in \cite{DAmbrosio:2020yfa,Christodoulou:2023psv}. In this work, we refine their technique and provide a more precise formulation, demonstrating that it is particularly effective for evaluating spin-summed amplitudes in the semiclassical regime.

\smallskip

The sums over the spins are concentrated around configurations satisfying the equation
\begin{equation}
\label{eq:LaplaceSpinsEq}
\phi_{ab} = \psi_{ab}(\ell_{ab}) \ ,
\end{equation}
which coincides with the equation we found in the classical theory \eqref{eq:eomref} and is solved by $\ell_{ab} = \lambda \, l_{ab}(\phi_{ab})$. We have already discussed the solutions to \eqref{eq:LaplaceSpinsEq}: there are infinitely many, labeled by a scale parameter $\lambda$. The solution fixes (up to scale) the lengths as functions of the dihedral angles $l_{ab}(\phi_{ab})$. Note that a solution to \eqref{eq:LaplaceSpinsEq} exists only if the relative orientation between the vertex and the spin network is the appropriate one. 
This was also derived explicitly—together with the Hessian of the Regge action, which coincides with ours—in \cite{Dittrich:2007wm}, where it was shown that the Hessian possesses a unique null direction associated with the scaling symmetry.
Only five out of the six spin sums can be evaluated directly. The remaining sum corresponds to the null direction of the Hessian, i.e., the overall scale. The final form of the amplitude takes the form:
\begin{equation}
\label{eq:final}
A_v(h_{ab})
\;\approx\;
\sum_{\lambda}\frac{\mathcal{N}(\lambda l_{ab})}{2\sqrt{-12 i \pi V(\lambda l_{ab})}} \sqrt{\frac{(2\pi \lambda^5)}{-\det H}} \exp\left({i \lambda \sum_{ab}l_{ab}(\phi_{ab}) \phi_{ab}}\right) + \text{inverted orientation} \ .
\end{equation}

\smallskip

The asymptotic form of the vertex amplitude \eqref{eq:final} explicitly shows that, in the semiclassical regime, the amplitude is dominated by classical tetrahedra compatible with the boundary holonomies. It oscillates with a frequency proportional to the Regge action of the tetrahedron.

The holonomies have been decomposed: the spinorial contribution is encapsulated in the intrinsic form factor $\mathcal{N}$, while the dihedral angles $\phi_{ab}$ govern the dynamics.

\smallskip

Note that for each tetrahedron at fixed scale $\lambda$, the edge lengths are completely determined by the dihedral angles (and thus by the holonomies), according to equation \eqref{eq:LaplaceSpinsEq}. Alternatively, they can be derived directly from the Hamilton principal function by taking its derivative with respect to the dual angle.

The appearance of an infinite number of contributions is not surprising, as the amplitude expressed in group variables can be exactly rewritten as a series of delta functions. When these delta functions are saturated, the result is necessarily divergent. Here, we have interpreted this divergence as being related to the scale invariance of an angle description of the vertex amplitudes.

\subsection{Classically Forbidden Transitions}

Finally, we are in a position to address the original question: what happens to the amplitude if there are no Euclidean tetrahedra compatible with the boundary holonomies? As we already discussed in the previous section, for a given set of angles such that the signature of the Gram matrix is $(2,1,1)$, there are only complex solutions to equation~\eqref{eq:eomref}. These solutions correspond to geometric tetrahedra that are not embeddable in Euclidean space, but can be embedded in Minkowski space. Therefore, if interpreted as the evolution of a boundary surface, the process is classically forbidden but becomes quantum mechanically allowed, as the analytic continuation of a classical geometry.

Hamilton's principal function for these geometries is obtained via analytic continuation of the Regge action for the tetrahedron, written in terms of the dihedral angles, as in equation~\eqref{eq:FOActionPhi}. These classically forbidden geometries dominate the amplitude, and the expression~\eqref{eq:final} is to be analytically continued to complex lengths. Each contribution at a fixed scale $\lambda$ is exponentially suppressed by the factor
\begin{equation}
\exp\left(- \lambda \sum_{ab} |l_{ab}(\phi_{ab})|\, \phi_{ab}\right) \ ,
\end{equation}
in full analogy with the analysis in the length representation, as originally hinted at in~\cite{Regge:1968} and more recently discussed in~\cite{Dona:2024rdq}.


\section{Discussion}
\label{sec:discussion}

Spinfoams provide a path integral formulation of Loop Quantum Gravity, describing quantum evolution of kinematical states associated with spatial slices. Classically, in three Euclidean dimensions, evolution corresponds to gluing tetrahedra to these slices. This evolution is governed by Hamilton's principal function, which is explicitly given by the Regge action evaluated on classical geometries \cite{Dittrich:2011ke}.

\smallskip

However, specific boundary data can be incompatible with classical evolution. In these situations, trajectories or geometries that are classically forbidden may dominate the path integral, though their contributions to the amplitude become exponentially suppressed. We explicitly illustrated this scenario within the context of 3D Euclidean quantum gravity described by the Ponzano-Regge spinfoam model. Although the interpretation in terms of length variables was already clear \cite{Regge:1968,Dona:2024rdq}, previously there was no concrete demonstration of how this corresponds precisely to the dual (holonomy) picture, nor how lengths emerge explicitly.

\smallskip

We showed that Hamilton's principal function, governing the classical theory, can be explicitly derived in the holonomy picture through a Legendre transform of the length formulation. Classically forbidden geometries appear naturally as analytic continuations of classically allowed geometries. In the length picture, lengths remain real, while the dual variables (dihedral angles) become purely imaginary, leading to geometries with negative squared volumes embeddable in Minkowski space.

In contrast, in the dual picture, boundary holonomies remain real $SU(2)$ elements, describing parallel transport between triangles of the spatial slice and preserving the reality of the dihedral angles, even for classically forbidden geometries. However, lengths become imaginary, corresponding to timelike edge lengths embedded in Minkowski space. In this representation, geometries again exhibit negative squared volumes, highlighting the duality and consistency between the length and holonomy formulations of spinfoam models.

In the length picture, tunneling trajectories correspond to Lorentzian tetrahedra with a spacelike boundary. In the angle picture, by contrast, we find Lorentzian tetrahedra with a timelike boundary, which are classically forbidden configurations. It would be interesting to investigate the conditions under which both lengths and dihedral angles become generically complex. Such situations should correspond to fully generic Lorentzian tetrahedra, potentially revealing a richer structure of quantum transitions.

\smallskip

This mechanism is what is likely to govern the black-to-white hole tunneling process in Covariant Loop Quantum Gravity \cite{Christodoulou:2016vny}, which provides one of the motivations for the present work. Calculations in the literature have relied heavily on coherent boundary states \cite{Christodoulou:2018ryl,Christodoulou:2023psv}. This approach has limitations when describing classically forbidden transitions, particularly within minimal discretizations. We have demonstrated that interpreting exponential suppression solely as tunneling events can be misleading, since a similar suppression naturally arises due to kinematical mismatches in boundary momenta. The black-to-white hole transition is characterized by its extrinsic curvature flip; the best way to characterize it is to focus on boundary states that are eigenstates of the extrinsic curvature.

\smallskip

Our analysis confirms that holonomies almost completely specify classical geometries, including classically forbidden ones, by fixing the shape of the tetrahedron but not its scale. As a consequence, the amplitude is dominated by an infinite number of classical geometries differing by an overall scale factor. At each fixed scale, classically forbidden configurations contribute an exponentially suppressed term proportional to the product of lengths and extrinsic curvature.
We also note that the need to understand the analytic continuation of the Ponzano–Regge amplitude in terms of complex dihedral angles was emphasized in \cite{Bonzom:2019dpg}, where the generating function for the $6j$ symbol is interpreted as a Ponzano–Regge amplitude with a scale-invariant boundary state defined solely by dihedral angles. This also leads to an infinite number of classical geometries differing by an overall scale factor, precisely as discussed in the present manuscript.

\smallskip

Extending these insights to four-dimensional Lorentzian models suggests that the suppression of tunneling amplitudes should take the form $\exp(-m^2\Phi)$, where $m$ relates to the black hole mass and $\Phi$ to the traversal time of the black-to-white hole spacetime. However, considerable additional work is required to completely validate this scenario.

\smallskip

We also identified a critical issue in prescribing boundary states solely by extrinsic curvature in the black-to-white transition. The scale ambiguity inherent in this formulation complicates interpretation, as the scale is directly tied to the black hole's mass, a physically significant parameter.

\smallskip

These challenges suggest that we may need to reassess the current calculation and refine the discretization strategy to ensure it accurately captures the essential physical features of the black-to-white hole transition. Specifically, future discretization schemes must account for: (1) the classically forbidden nature of the transition region; (2) the relationship between the intrinsic geometry of the discretization and the black hole mass; and (3) the mirrored intrinsic geometries and flipped extrinsic curvatures of black and white hole regions.

Addressing these considerations will be crucial for future theoretical and numerical advances in quantum gravitational descriptions of black hole transitions. A possible solution could be to consider a hybrid boundary state. We might construct a boundary state that is an eigenstate of the length operator (area in 4D) for the perimeter links, and an eigenstate of the holonomy (extrinsic curvature) for the remaining links. Such a hybrid state could naturally fix the scale, perhaps directly in terms of the physical parameters of the black hole, and simultaneously characterize the spatial surfaces of both the black and white holes purely through their extrinsic curvature. This approach would allow the use of necessary symmetries. However, this proposal remains speculative, and a detailed analysis, possibly involving a comprehensive rethinking of the triangulation scheme, is left for future work.

\smallskip

Additionally, it would be interesting to extend our analysis to Lorentzian models. We limited our study to the Euclidean model for simplicity. However, several Lorentzian (2+1) spinfoam models exist \cite{Davids:2000kz, Freidel:2000uq, Freidel:2005bb,Borissova:2024pfq,Simao:2024don}. Investigating how our results generalize to these cases would be beneficial and is expected to yield similar insights.

\smallskip

In this context, a natural extension is the possible connection to the BTZ black hole \cite{Banados:1992wn}. Could similar techniques be employed to study a quantum version of the BTZ black hole \cite{Geiller:2013pya}? Specifically, does it also tunnel into a BTZ white hole? Notably, this scenario requires a cosmological constant, suggesting the need to explore quantum group formulations of spinfoam theory \cite{Haggard:2015ima, Haggard:2015yda, Han:2021tzw, Han:2025mkc}. A positive aspect here is that curved geometry naturally fixes the scale due to the inherent cosmological length scale.

\smallskip

Finally, similar methods could be utilized in studying bouncing cosmology within spinfoam cosmology frameworks \cite{Bianchi:2010zs, Vidotto:2011qa}. Our insights might help characterize the spinfoam equivalent of the Hartle–Hawking state, potentially leading to interesting cosmological applications of the theory.

\section{Acknowledgments}
This work is supported by the ID\# 62312 grant from the John Templeton Foundation, as part of the ``The Quantum Information Structure of Spacetime (QISS)'' Project (\href{qiss.fr}{qiss.fr}).  H.M.H. and C.R. acknowledge support from the Perimeter Institute for Theoretical
Physics through its affiliation program and its distinguished research chair program, respectively. Research at Perimeter Institute
is supported by the Government of Canada through the Department of Innovation, Science, and Economic
Development, and by the Province of Ontario through the Ministry of Colleges and Universities.


\begin{appendices}

\section{Dihedral angles giving rise to tetrahedra and the Gram matrix}
\label{app:Gram}

The geometric interpretation of the Gram matrix and its structure provides a useful technique for reconstructing simplices from angular data. In particular, when the Gram matrix satisfies a specific set of algebraic properties, it encodes the complete geometry of an $n$-simplex. This appendix makes these connections precise. We focus on zero-curvature simplices, though many of the useful properties of Gram matrices generalize to simplices of non-zero constant curvature \cite{Haggard2016,feng_luo}. The utility of Gram matrices, using a variety of formulations, has been a recurring theme in the quantum gravity literature, e.g.  \cite{Freidel:2002mj,Dittrich:2008va,Bonzom:2011jh}.

 For an $n$-simplex embedded in a space of Euclidean signature, the Gram matrix $G$ is defined by the cosines of the dihedral angles between its $(n{-}1)$-dimensional facets:
\begin{equation}
\label{eq:AppGram}
    G_{ij} = \cos\phi_{ij}, \quad \text{with} \quad \phi_{ii} = 0,
\end{equation}
where $\phi_{ij}$ is the external angle between the $i$-th and $j$-th facets. More generally, the precise form of the Gram matrix depends on the metric signature of the embedding space, but the closure condition
\begin{equation}
    \sum_{i=1}^{n+1} A_i \, \hat{n}_i = 0
\end{equation}
imposes a universal geometric constraint. Here $A_i$ is the $(n-1)$-volume (e.g., area for $n=3$, volume for $n=4$) of the $i$-th facet, and $\vec{n}_i$ is its unit outward-pointing normal vector.

As a consequence, the Gram matrix is always singular, with rank $n$, for a closed simplex. Its one-dimensional kernel corresponds precisely to the vector of facet volumes. In other words, the kernel of $G$ is spanned by a single \emph{positive} vector
\begin{equation}
    A = (A_1, A_2, \ldots, A_{n+1}),
\end{equation}
encoding the magnitude of the facets, for example, in 3D, the areas for a tetrahedron. This geometric interpretation holds regardless of the metric signature of the space (e.g., Euclidean, Lorentzian). Hence, the closure condition and corresponding rank deficiency of the Gram matrix provide a signature-independent way of characterizing simplices.

To make this precise, consider a real, symmetric matrix of size $(n{+}1) \times (n{+}1)$ whose diagonal entries are $1$ and such that:
\begin{enumerate}[label=A\arabic*, ref=A\arabic*]
    \item The matrix is singular with rank $n$. \label{assumption1}
    \item There exists a \emph{positive} vector $A = (A_1, \ldots, A_{n+1})$ such that $G A = 0$.\label{assumption2}
\end{enumerate}
Because $G$ is real and symmetric, it is diagonalizable by an orthogonal matrix $Q$,
\begin{equation}
\label{eq:symdiag}
    G = Q \tilde{G} Q^\top,
\end{equation}
with $\tilde{G}$ the diagonal matrix of eigenvalues of $G$. The condition $\mathrm{rank}(G) = n$ implies that one of these eigenvalues is zero, reflecting the fact that the simplex satisfies closure.

A real, symmetric matrix $G$ that satisfies the conditions described above can be taken to be the Gram matrix of a geometrical $n$-simplex. However, for this simplex to be embeddable in a metric space, the matrix must admit a factorization consistent with the metric of the embedding space. This is formalized by the condition that $G$ can also be written as
\begin{equation}
\label{eq:diag}
G = P^\top \eta P,
\end{equation}
where $\eta$ is the metric tensor of the ambient space and $P$ is a matrix whose columns represent the normal vectors to the facets of the simplex, appropriately normalized. This condition is what properly generalizes Equation~\eqref{eq:AppGram} to arbitrary signature and guarantees that the simplex can be embedded in a space with the appropriate Euclidean or Lorentzian signature. For $n>3$, general metric signatures are possible, though with less clear physical interpretations. The proof below is equivalent to the singular value decomposition proofs of \cite{HaggardLittlejohn2010,Haggard2011}, and, in this context, was inspired by the work \cite{Asante:2025qbr}.

\paragraph{Euclidean case.} A tetrahedron is embeddable in Euclidean space if its Gram matrix $G$ is positive semi-definite with signature $(n,0,1)$, that is, with $n$ positive eigenvalues and one zero eigenvalue corresponding to closure. In this case, the diagonalized form of the Gram matrix, the $\tilde{G}$ of \eqref{eq:symdiag}, has non-negative entries with a single zero. Because $\tilde{G}$ is positive semi-definite, we can define a square root matrix $\tilde{G}^{1/2}$, and rewrite the diagonalization as
\begin{equation}
\label{eq:diag2}
G = Q \tilde{G}^{1/2} \eta \tilde{G}^{1/2} Q^\top,
\end{equation}
with $\eta = \mathbb{I}$ the Euclidean metric. Identifying $P = \tilde{G}^{1/2} Q^\top$, this confirms that $G$ admits the required factorization \eqref{eq:diag}, and thus corresponds to a Euclidean simplex. (Of course, here and below, the last row of zeroes in $P$ can be dropped to recover the standard form of the normal vectors.)

\paragraph{Lorentzian case.} In contrast, a tetrahedron is embeddable in Minkowski space if the Gram matrix has signature $(n-1,1,1)$: $n-1$ positive eigenvalues, one negative, and one zero. While there is no general constructive theorem that guarantees a decomposition of the form \eqref{eq:diag} in the Lorentzian case, the same diagonalization of $G$ as a symmetric matrix \eqref{eq:symdiag} still holds. Let us order the eigenvalues in $\tilde{G}$ such that the first entry is the negative one. We can then define a matrix $P$ such that
\begin{equation}
P = |\tilde{G}|^{1/2} Q^\top,
\end{equation}
and use the identity $|\tilde{G}|^{1/2} \eta |\tilde{G}|^{1/2} = \tilde{G}$, where $\eta = \mathrm{diag}(-1,1,\dots,1,0)$ is the Minkowski metric in a basis adapted to the eigenvalue signature. This leads to the identity
\begin{equation}
P^\top \eta P = Q \tilde{G} Q^\top = G,
\end{equation}
thus establishing that $G$ corresponds to a realizable tetrahedron in Minkowski space.

This decomposition provides a clear and coordinate-independent criterion: the signature of the Gram matrix determines the embeddability of the simplex, and whether the corresponding geometry is Euclidean or Lorentzian. Assumptions \ref{assumption1} and \ref{assumption2} have another important consequence for the realizability of given data: they determine the structure of the cofactors of a related area Gram matrix, and this, in turn, can be used to fix the scale of the embedded simplex.

\paragraph{Gram matrix cofactors.} Given a matrix $M$, its $ij$-th minor, denoted $\hat{M}_{ij}$, is the determinant of the matrix obtained by removing the $i$-th row and $j$-th column of $M$. The $ij$-th cofactor is this minor adjusted by a sign, $\hat{C}_{ij} = (-1)^{i+j} \hat{M}_{ij}$. Define a matrix $\mathsf{A} = \textrm{diag}(A_1, \dots, A_{n+1})$, and a new area Gram matrix by
\begin{equation}
G_A = \mathsf{A} G \mathsf{A}^\top = \mathsf{A} G \mathsf{A}. 
\end{equation}
The cofactors of the area Gram matrix have a special structure. 

Consider the sum of the entries of the $i$-th row of $G_A$. Let $\vec{A}_i = A_i \hat{n}_i$, then this sum is given by 
\begin{equation}
    \vec{A}_i \cdot \sum_j \vec{A}_j = A_i \hat{n}_i^\top \eta \Big(\sum_j A_j \hat{n}_j \Big) =0 \qquad \textrm{(fixed $i$)},
\end{equation} 
where, again, $\eta$ is the metric of the embedding space; the sum vanishes due to closure. Because $G_A$ is symmetric, this also holds for all of the columns. In mathematics, a square matrix all of whose rows and columns sum to the same value is called semi-magic.\footnote{If the diagonal and anti-diagonal also sum to the same value, the matrix is termed magic.} Hence, $G_A$ is semi-magic.  The importance of this is that semi-magic matrices with vanishing row and column sums have all cofactors identically equal. For a proof, see \cite{MathStackExWeb}, but the physical implication is immediate: each cofactor of $G_A$ is simply a different way of computing the volume $V$ of the $n$-simplex to a power, specifically  $V^{2(n-1)}$. For $G_A$, and hence $G$, to correspond to a geometrical $n$-simplex it is essential that all of these ways of computing the volume must agree.  The specific power $2(n-1)$ can be proven by induction or by dimensional analysis. Here we forgo the induction, choosing instead to illustrate the simplest non-trivial case in the hope that that will best serve the reader's intuition. 


Given the normal vectors $\hat{n}_i$ to the triangles of the tetrahedron and the null vector of areas $(A_1, \dots, A_4)$, we can construct the area Gram matrix $G_A$ out of the matrix whose columns are the area vectors $\vec{A}_i = A_i \hat{n}_i$ and its transpose
\begin{equation}
G_A = \begin{bmatrix}
\vec{A}_1 \\
\vec{A}_2 \\
\vec{A}_3 \\
\vec{A}_4 
\end{bmatrix} . 
\begin{bmatrix}
\vec{A}_1 & \vec{A}_2 & \vec{A}_3 & \vec{A}_4 
\end{bmatrix} \ .
\end{equation}
The $ij$-th area cofactor, call it $\hat{C}_{ij}$, is obtained by removing the $i$-th entry from the row matrix and the $j$-th entry from the column matrix. As an example,
\begin{align}
\hat{C}_{12} &= (-1)^{1+2} 
\left \lvert 
\begin{bmatrix}
\vec{A}_2 \\
\vec{A}_3 \\
\vec{A}_{4}
\end{bmatrix} . 
\begin{bmatrix}
\vec{A}_1 &
\vec{A}_3 &
\vec{A}_{4}
\end{bmatrix} 
\right \rvert = (-1)^{3} 
\left \lvert 
\begin{bmatrix}
\vec{A}_2 \\
\vec{A}_3 \\
\vec{A}_{4}
\end{bmatrix}
\right \rvert . 
\left \lvert 
\begin{bmatrix}
\vec{A}_1 &
\vec{A}_3 &
\vec{A}_{4}
\end{bmatrix} 
\right \rvert \ ,
\label{eq:cofac}
\end{align}
here the determinant function distributes over the $3 \times 3$ matrices of the vectors that were not omitted. But, the determinant of the matrix of 3 area vectors of a tetrahedron has a geometrical meaning
\begin{equation}
\left \lvert 
\begin{bmatrix}
\vec{A}_i &
\vec{A}_j &
\vec{A}_k
\end{bmatrix} 
\right \rvert = \vec{A}_i \cdot (\vec{A}_j \times \vec{A}_k) = \dfrac{9}{2} V^2\ ,
\label{eq:determ}
\end{equation}
where $V^2$ is the (signed) squared volume of the tetrahedron. The  cofactor sign ensures that all of the triple products are computed with the same orientation and so all cofactors give strictly the same positive result
\begin{align}
\hat{C}_{ij} = \dfrac{81}{4} V^4\ .
\label{eq:cofactors}
\end{align}
This illustrates the general result of the previous paragraph. 

\paragraph{A first method for fixing the physical scale and allowed dihedral angles.}
The results of the last subsection have two important consequences: they provide a natural way to fix the physical scale in the angle description in this paper, and they illustrate an important consistency check needed when taking the angle approach. We discuss each of these consequences in turn. 

As was emphasized in Section~\ref{sec:tetra}, the description of zero-curvature tetrahedra in terms of dihedral angles is necessarily scale invariant. A particular physical scenario will usually fix this overall scale, and it is important to have a mechanism to fix the scale in the model. The result above on cofactors furnishes a simple way to do this if the physical scenario fixes the volume of the model building blocks, i.e., the tetrahedra in the Ponzano-Regge model. Concretely, this can be achieved by the following algorithm: Specify your angles, e.g., along the lines of the example in Section~\ref{sec:tetra}, and compute the null vector of the resulting angle Gram matrix. If the entries of the null vector are not all positive, this is not a good angle configuration (this is the criterion that leads to no constructible tetrahedra below the diagonal in Figure~\ref{fig:transition-region}). If they are positive, then use them to form the area Gram matrix and compute any cofactor. Scale this cofactor to reach the desired physical volume scale.  Because the cofactor involves $V^4$, this uniformly works for both classically allowed and forbidden configurations with the same sign. 

This short algorithm highlights another feature of the cofactor result: by selecting only angle Gram matrices with completely positive null vectors, we are picking out the physical dihedral angles that lead to constructible tetrahedra, both Euclidean and Lorentzian. Indeed, in \cite{feng_luo}, Feng Luo proves that dihedral angles leading to a Euclidean simplex must satisfy three conditions on their angle Gram matrix: (a) $\det G = 0$, (b) all angle cofactors $\hat{G}_{ij}$ must be positive, and (c) all principal submatrices of $G$ must be positive definite. Principal submatrices are those that arise by deleting the same row and column, and an effective way of determining whether one is positive definite is by checking that all of its principal minors are positive. Our assumption \ref{assumption1} agrees with (a), and it is now apparent that our assumption \ref{assumption2} is a more physical manifestation of Luo's (b).  

There is a good reason that we have no condition corresponding to Luo's condition (c). Relaxing this condition is exactly what allows for the classically forbidden Lorentzian tetrahedra. Nonetheless, this condition plays an interesting role in our analysis too. The curious reader will find several efforts in the literature to directly characterize the angles giving rise to Euclidean tetrahedra via inequalities on these angles and their cosines. The angle Gram matrix conditions above are by far the most efficient way of doing this. However, through a bit of experimentation, one will quickly find that these conditions are often redundant if all you want is to determine which regions in the angle space correspond to Euclidean tetrahedra. With our richer dynamical perspective, however, the vanishing of any principal minor carries interesting information, often signalling a transition between different kinds of Lorentzian tetrahedra as in the example at the end of Section~\ref{sec:tetra}. 

The angle Gram matrix is the essential tool for characterizing the space of geometrical tetrahedra in terms of dihedral angles.

\paragraph{A second method: the circumradius as scale factor.}

Many of the computations familiar from length Regge calculus can also be done completely in the angle formulation. Here we lay out these formulas and explore a second method of setting the physical scale using the circumradius of a tetrahedron. 

Following \cite{Asante2018degrees} at the outset, we use Jacobi's formula for the derivative of the determinant of the Gram matrix to find
\begin{equation}
\dfrac{\d \det G}{\d \phi_{ij}} = - \sin{\phi_{ij}} \hat{G}_{ij} \ ,
\label{eq:Jacobi}
\end{equation}
where $\hat{G}_{ij}$ is the $ij$-th cofactor of the angle Gram matrix. From Equations~\eqref{eq:determ} and \eqref{eq:cofactors} we get 
\begin{equation}
    \hat{G}_{ij} = \dfrac{81}{4} \dfrac{V^4}{(A_1 A_2 A_3 A_4)^2} A_i A_j 
    \label{eq:dev_cof}
\end{equation}
and  substituting this into  \eqref{eq:Jacobi}, we get
\begin{equation}
\dfrac{\d \det G}{\d \phi_{ij}} = - \dfrac{81}{4} \dfrac{V^4}{(A_1 A_2 A_3 A_4)^2} A_i A_j \sin{\phi_{ij}} = - \dfrac{81}{4} \dfrac{V^5}{(A_1 A_2 A_3 A_4)^2} \dfrac{A_i A_j \sin{\phi_{ij}}}{V} = - \dfrac{3^5}{2^3} \dfrac{V^5}{(A_1 A_2 A_3 A_4)^2} l_{ij} \ ,
\label{eq:Jacobi_final}
\end{equation}
where $l_{ij}$ is the edge length around which the dihedral angle $\phi_{ij}$ is measured. 

We can rewrite this relation using a scale unique to a tetrahedron. One convenient, and seemingly embedding independent choice (if we limit ourselves to real dihedral angles) is the tetrahedron's circumradius $R$.  From \cite{tetra_trig} and \cite{Cho:2000}, we have 
\begin{equation}
\dfrac{R^2}{-\det M} = \dfrac{2^2}{3^{10}} \dfrac{(A_1 A_2 A_3 A_4)^4}{V^{10}} \ , \quad \text{and} \quad (A_1 A_2 A_3 A_4)^3 = \dfrac{(6V)^8}{8^4(\hat{G}_{11} \hat{G}_{22} \hat{G}_{33} \hat{G}_{44})^{1/2}} \ ,
\label{eq:geom_formulas}
\end{equation}
where $R$ is the circumradius of the tetrahedron and $M$ is related to the Hadamard square of the Gram matrix and is defined by 
\begin{equation}
M_{ij} = \sin^2 \phi_{ij}  = 1 - G_{ij}^2\ , \quad \text{with} \quad \phi_{ii}=0 \ .
\label{eq:Mmatrix}
\end{equation}
Therefore,  
\begin{equation}
\dfrac{\d \det G }{\d \phi_{ij}}  = - \dfrac{1}{4 R} \sqrt{-\det M} \, l_{ij} \ .
\label{eq:Gram_var}
\end{equation}
Using the above formulas, we derive the squared edge lengths in terms of the circumradius
\begin{equation}
l_{ij}^2 = \dfrac{16 R^2}{-\det M} \sin^2{\phi_{ij}} \cdot \dfrac{\hat{G}_{ij}}{\hat{G}_{kl}} \cdot \sqrt{\hat{G}_{11} \hat{G}_{22} \hat{G}_{33} \hat{G}_{44}} \ ,
\label{eq:lengths2}
\end{equation}
where $kl$ denotes the pair of indices disjoint to $ij$, e.g. if $ij=12$, then $kl=34$.  We observe that classically forbidden tetrahedra, when parametrized in terms of real dihedral angles, are such that $R > 0$, $\hat{G}_{ij}>0$ and $\det M > 0$. Therefore, $l_{ij}$ is imaginary, and all the edges of the tetrahedron are timelike.

Similarly, the volume of the tetrahedron expressed in terms of the dihedral angles and the circumradius $R$ is given by \cite{Cho:2000}

\begin{equation}
V = \dfrac{32}{3} \cdot \dfrac{\hat{G}_{11} \hat{G}_{22} \hat{G}_{33} \hat{G}_{44}}{(-\det M)^{3/2}} \cdot R^3 \ .
\label{eq:volume}
\end{equation}
At first glance, the formula for the lengths \eqref{eq:lengths2} looks quite distinct from that derived from the Regge action \eqref{eq:eomangles}. However, using these results, we can show that \eqref{eq:lengthform} is equivalent to \eqref{eq:lengths2}:
\begin{align}
l_{ij}^2 &= \dfrac{16 R^2}{-\det M} \sin^2{\phi_{ij}} \cdot \dfrac{\hat{G}_{ij}}{\hat{G}_{kl}} \cdot \sqrt{\hat{G}_{11} \hat{G}_{22} \hat{G}_{33} \hat{G}_{44}} \nonumber \\
& = \dfrac{16 R^2}{-\det M} \sin^2{\phi_{ij}} \cdot \dfrac{A_i A_j}{A_k A_l} \cdot \sqrt{\hat{G}_{11} \hat{G}_{22} \hat{G}_{33} \hat{G}_{44}} \nonumber \\
& = \left( \dfrac{2}{3} \right)^2 \left(  \dfrac{A_i A_j}{V} \right )^2 \sin^2{\phi_{ij}} \ ,
\end{align}
where we used \eqref{eq:dev_cof} in the second equality and applied \eqref{eq:geom_formulas} to derive the result.


\section{Spin networks in the semiclassical regime}
\label{app:semiclassicalspinnetwork}
In this appendix, we present a detailed derivation of the asymptotic expression for the spin network wavefunction of Equation~\eqref{eq:spin_network}
\begin{equation}
    \label{eq:spin network_app}
    \Psi_{v}^{j_{ab}}(h_{ab})=\bigotimes_{n} i_{n} \bigotimes_{(ab)} D^{j_{ab}}(h_{ab}) \ ,
\end{equation}
here we omit the magnetic indices, which are contracted and summed according to the connectivity of the graph in Figure~\ref{fig:spinnetwork}. The intertwiner $i_a$ of a three-valent node $a$, and acting as the source of three links, is
\begin{equation}
\ket{i_a} = i^{m_{ab} m_{ac} m_{ad}} \ket{j_{ab} \, m_{ab} }\ket{j_{ac} \, m_{ac} }\ket{j_{ad} \, m_{ad} }  \ , \quad \text{with} \quad i^{m_{ab} m_{ac} m_{ad}} = \begin{pmatrix}
    j_{ab} & j_{ac} & j_{ad} \\
    m_{ab} & m_{ac} & m_{ad}
\end{pmatrix} \ ,
\label{eq:3valentinter}
\end{equation}
here $i^{m_{ab} m_{ac} m_{ad}}$ is the Wigner $3j$-symbol and $b,c,d\neq a$. Similarly, the Wigner matrix 
\begin{equation}
D^{j_{ab}}_{n_{ab}m_{ab}}(h_{ab}) = \bra{j_{ab},n_{ab}} D^{j_{ab}}(h_{ab}) \ket{j_{ab},m_{ab}} \ ,
\end{equation}
represents the SU(2) holonomy associated with the group variable $h_{ab}$. 
We indicate a spinor and its conjugate transpose by $\ket{z}$ and $\bra{z}$, with
\begin{equation}
\ket{z} := \left(\begin{array}{c}
z_{0}\\
z_{1}
\end{array}\right)
\ ,
\qquad 
\bra{z} := \left(
\bar{z}_{0},\  \bar{z}_{1}
\right)
\ .
\end{equation}
This Dirac notation simplifies the bookkeeping of spinorial indices. Spinor space is equipped with a natural inner product given by $\braket{w}{z}:= \bar{w}_0 z_0 + \bar{w}_1 z_1$ and a duality map $\mathcal{J}: \C^2 \to \C^2$, given by 
\begin{equation}
	\mathcal{J} \ket{z} = \sket{z} := \left(\begin{array}{c}
-\bar{z}_{1}\\
\bar{z}_{0}
\end{array}\right)\ .
\end{equation}
Moreover, a spinor and its dual form an orthogonal basis of $\C^2$ and represent a vector in $\R^3$ built from the matrix elements of the Pauli matrices $
\bra{z}\vec{\sigma} \ket{z} = -\vec n$ and  $
\sbra{z}\vec{\sigma} \sket{z} = \vec n$. We work with unit norm spinors to simplify calculations (and without loss of generality). 
The degrees of freedom of the holonomy $h_{ab}$ can be effectively parametrized using the twisted geometry parametrization introduced in \cite{Freidel:2010tt,Borja:2010rc,Livine:2011gp} and recently used to describe the emergence of geometries from locally flat spin foam theories \cite{Dona:2022hgr}. This parametrization reads
\begin{equation}
  h_{ab}
  \;=\;
  e^{i\frac{\phi_{ab}}{2}}\,\sket{z_{ba}}\bra{z_{ab}}
  \;-\;
  e^{-i\frac{\phi_{ab}}{2}}\,
     \ket{z_{ba}}\sbra{z_{ab}}\ .
\label{eq:holonomy2}
\end{equation}
The spinors represent the normals to the sides of the triangles within their respective planes, thereby encoding the intrinsic geometry (i.e., the edge shapes) of the source and target triangles of the holonomy. The angle $\phi_{ab}$ encodes the rotation between the source and target, capturing the extrinsic curvature between the source and the target triangles.

\smallskip

To derive the asymptotic expression for \eqref{eq:spin network_app}, we replace the resolution of the identity over the orthonormal basis $ \ket{j_{ab} \, m_{ab} }$ with the one over the overcomplete spinorial basis
\begin{equation}
    \sum_{m_{ab}} \ket{j_{ab} \, m_{ab} }\bra{j_{ab} \, m_{ab} }  \quad \longrightarrow \quad \ \int \d \mu(w_{ab}) \, \ket{j_{ab} \, w_{ab} }\bra{j_{ab} \, w_{ab} } \ ,
\end{equation}
where $\d \mu(w_{ab})$ is the normalized Gaussian measure on $\mathbb{C}^2$ \cite{Livine:2011gp}. 
In terms of the spinorial basis, the Wigner matrix becomes exponential
\begin{equation}
\label{WigExp}
\bra{j_{ab},w_{ba}} D^{j_{ab}}(h_{ab}) \ket{j_{ab},w_{ab}} = \bra{w_{ba}} h_{ab} \ket{w_{ab}}^{2j_{ab}} = e^{2j_{ab}\log \bra{w_{ba}} h_{ab} \ket{w_{ab}}} \ ,
\end{equation}
while the intertwiner becomes a coherent intertwiner sharply peaked around the classical geometry of a triangle \cite{Livine:2007vk}: $j_{ab}$ describes the length of edges of the triangle while the spinors $w_{ab}$ describe the unit normals $\hat{n}_i$ to these edges in the plane of the triangle. The norm of the coherent intertwiner, $c_a(w_{ab})$, is given by
\begin{equation}
c_a(w_{ab})=\bra{i_a} \Big(\ket{j_{ab}\, w_{ab}}\ket{j_{ac}\, w_{ac}}\ket{j_{ad}\, w_{ad}}\Big)= \bra{i_a} \int_{SU(2)} \! \! \! dg \ g \triangleright \ket{j_{ab}\, w_{ab}}\ket{j_{ac}\, w_{ac}}\ket{j_{ad}\, w_{ad}} \ ,
\end{equation}
and is exponentially suppressed unless the normals to the edges satisfy the closure condition \cite{Livine:2007vk}. Geometrically, this means that the spinors must be compatible with the intrinsic geometry (shape) of the triangle determined by the spins, or equivalently, by the edge lengths.

We perform the spinorial integration using the complex Laplace method (saddle-point method). We focus solely on the integral of the Wigner matrix, which we assume dominates. The coherent intertwiner also contains an oscillatory part, however, to simplify the calculation, we neglect it and assess its impact a posteriori. 

For each link, there are two integrals: one over $w_{ab}$ and one over $w_{ba}$. We represent the contributions from the nodes using a generic function, $f(w_{ab},w_{ba})$ as a placeholder:

\begin{equation}
\int \d\mu(w_{ab})\d\mu(w_{ba})\, f(w_{ab},w_{ba})\, e^{2j_{ab}\log \bra{w_{ba}} h_{ab} \ket{w_{ab}}} \ .
\label{eq:ampliapprox1}
\end{equation}
We parametrize the holonomy as in \eqref{eq:holonomy2}, and take it to be fixed.
The stationary phase equations are
\begin{equation}
h_{ab} \ket{w_{ab}}  = e^{\frac{i \alpha}{2}} \ket{w_{ba}} \ , \quad h_{ab} \sket{w_{ab}}  = e^{-\frac{i \alpha}{2}} \sket{w_{ba}} \ .
\label{eq:stationeqs}
\end{equation}
These have two possible solutions
\begin{equation}
\ket{w_{ab}} = e^{i \beta} \ket{z_{ab}} \ \text{and}\ \sket{w_{ba}} =  e^{i \beta'} \ket{z_{ba}} \ , \quad  \text{or} \quad \sket{w_{ab}} = e^{i \chi} \ket{z_{ab}} \ \text{and}\ \ket{w_{ba}} =  e^{i \chi'} \ket{z_{ba}} \ .
\label{eq:stationeqs_imp}
\end{equation}
The integral \eqref{eq:ampliapprox1} is peaked when the spinors $w$ are aligned or anti-aligned with the source and target spinors associated with the SU(2) holonomy $h_{ab}$ assigned to the link $(ab)$. The phases $\beta$ and $\chi$ remain arbitrary and are completely redundant due to the invariance of the original expression. We fix these phase factors to $1$. As a result, the integral extracts the extrinsic geometry (specifically, the dihedral angle $\phi_{ab}$) from the holonomy and encodes it as a phase

\begin{equation}
\label{eq:limit_oneterm}
\int \d\mu(w_{ab})\d\mu(w_{ba})\, f(w_{ab},w_{ba})\, e^{2j_{ab}\log \bra{w_{ba}} h_{ab} \ket{w_{ab}}} \approx f(z_{ab},z_{ba})\dfrac{(2 \pi)^2}{\sqrt{|H_S(z)|}}e^{-i j_{ab} \phi_{ab}} + \text{inverted orientation} \ .
\end{equation}

The Hessian, $|H_S(z)|$, can also be computed at the stationary points \eqref{eq:stationeqs}. From \eqref{WigExp} the exponent is
\begin{equation}
S_{st} =  2j_{ab}\log \bra{w_{ba}} h_{ab} \ket{w_{ab}} := 2 j_{ab} \log T \ .
\label{eq:exponent}
\end{equation}
Thus, the first and second variations are
\begin{equation}
\delta S_{st} = 2 j_{ab} \:  \dfrac{1}{T} \delta T \:\:\:\: \text{and} \:\:\: \delta^2 S_{st} = 2 j_{ab} \left( \dfrac{1}{T} \delta^2 T - \dfrac{(\delta T)^2}{T^2} \right) \ .
\label{eq:hess}
\end{equation}
In our case, the spinors vary in such a way that they preserve their (unit) norms, and so  
\begin{equation}
\ket{w_{ab}} \longrightarrow \dfrac{\ket{w_{ab}} + \epsilon \ket{v_{ab}}}{|| \: \ket{w_{ab}} + \epsilon \ket{v_{ab}} \: ||} \approx \ket{w_{ab}} + \epsilon \ket{v_{ab}} - \frac{1}{2}\epsilon^2 (\braket{v_{ab}}{v_{ab}}) \ket{w_{ab}} + O(\epsilon^3) \ .
\label{eq:spinor_vari}
\end{equation}
and similarly for $\ket{w_{ba}}$. 
Plugging in $T=\bra{w_{ba}} h_{ab} \ket{w_{ab}}$ and the variation of the spinors $\ket{w_{ab}}$ and $\ket{w_{ba}}$ and collecting terms with one and two powers of $\epsilon$, we get 
\begin{equation}
    \delta T = \bra{w_{ba}} h_{ab} \ket{v_{ab}} + \bra{v_{ba}} h_{ab} \ket{w_{ab}}  \ , \quad \text{and} \quad \delta^2 T = 2 \bra{v_{ba}} h_{ab} \ket{v_{ab}} - \bra{w_{ba}} h_{ab} \ket{w_{ab}} \left(\braket{v_{ab}}{v_{ab}} +\braket{v_{ba}}{v_{ba}} \right) \ .
\end{equation}
The Hessian Matrix of the function $S_{st}$ \eqref{eq:exponent} can then be derived from \eqref{eq:hess} decomposing the perturbation spinors $v_{ab}$ and $v_{ba}$ in the respective $w_{ab}$ and $w_{ba}$ spinorial basis. We can write the Hessian in the basis $(\ket{w_{ab}}, \sket{w_{ab}},\ket{w_{ba}}, \sket{w_{ba}})$ and get:
\begin{equation}
H_{S} =  \begin{pmatrix}
0 & 0 & - j_{ab} & 2 j_{ab}\frac{\sbra{w_{ba}} h_{ab} \sket{w_{ab}}}{\bra{w_{ba}} h_{ab} \ket{w_{ab}}} \\
0 & 0 & 0 & - j_{ab} \\
- j_{ab} & 0 & 0 & 0 \\
 2 j_{ab} \frac{\sbra{w_{ba}} h_{ab} \sket{w_{ab}}}{\bra{w_{ba}} h_{ab} \ket{w_{ab}}} & - j_{ab} & 0 & 0
\end{pmatrix} \ .
\label{eq:HessianMat}
\end{equation}
The Hessian evaluated at the stationary point(s) is thus
\begin{equation}
|H_S| = (j_{ab})^4 \ ,
\label{eq:Hessian}
\end{equation}
or at the same level of approximation that we are working in the large scaling limit,  $(\ell_{ab})^4$. If we now reassemble the six Wigner matrices on the links of our spin network with the four coherent state norms (coming from the nodes), we observe that each function $c_a(w_{ab})$ gets evaluated at the spinors associated with the holonomies, denoted $z_{ab}$. Notice that the closure condition imposed by the intertwiners forces the saddle points found in the link integrals to be synchronized. Therefore, there are only two contributing saddle points altogether.

Moreover, the asymptotic form of $c_a(z_{ab})$ is known, as it was explicitly computed in \cite{Livine:2007vk}, and it is non-oscillatory provided the spinors are compatible with the spins. The assumption of computing the link integrals independently is justified: the only additional constraint introduced by the intertwiners is the closure condition, which is independent of the equations governing the link integrals.

Thus the asymptotic expansion of the spin network wavefunction is 
\begin{equation}
\Psi_{v}(h_{ab})^{j_{ab}} \approx \mathcal{N}(\ell_{ab})e^{-i \sum_{ab}\ell_{ab} \phi_{ab}} + \text{inverted orientation} \ ,
\end{equation}
where the multiplicative factor is
\begin{equation}
\mathcal{N}(j_{ab}) =  (2 \pi)^{12} \dfrac{\prod_n c_n(z_{ab})}{\prod_{ab} (\ell_{ab})^2} \ .
\label{eq:mult_fact}
\end{equation}

\section{The discrete Laplace method}
\label{app:Laplace}
The Laplace method is a technique for approximating integrals that depend on a large-scale parameter \cite{deBruijn:1981,BenderOrszag:1999}. It is widely used in physics and, in its complex, steepest-descent form, in spin-foam calculations \cite{Barrett:2009mw,Han:2011rf,Han:2011re,Engle:2015zqa,Bahr:2015gxa,Dona:2017dvf,Dona:2019dkf,Dona:2020yao}. The same idea can also be applied to derive the asymptotic expansion of sums. We briefly review this version of the method here. Suppose we wish to evaluate sums of the form
\begin{equation}
  S = \sum_{k=a}^b g(k)\,e^{f(k)} \ ,
\end{equation}
where $f$ is assumed to be a smooth, homogeneous function, attaining its unique global maximum at some point $k_0$ in the interval $a<k_0<b$, and $g(k)$ varies slowly in comparison to the rapid exponential growth induced by $e^{f(k)}$. Introducing a large parameter $\lambda\gg1$ and using homogeneity of $f$, we can write
\begin{equation}
  S = \sum_{k=a}^b g(k)\,\exp\bigl(\lambda f(k/\lambda)\bigr) \ .
\end{equation}
Intuitively, as $\lambda\to\infty$, terms in the sum for which $x=k/\lambda$ lies far from $x_0=k_0/\lambda$ become exponentially small, so the main contribution arises from $k$ near to $k_0$. We assume $x_a=a/\lambda < x_0 < x_b=b/\lambda$. By expanding both the exponent and the prefactor in a neighborhood of $x_0$, one replaces the discrete sum by a Gaussian-type sum, which in turn is approximated by a Gaussian integral.

In practice, the procedure unfolds in three stages. First, one locates the saddle point by solving
\begin{equation}
  f'(x_0)=0 \ ,\quad f''(x_0)<0 \ ,
\end{equation}
confirming a local maximum. Next, Taylor expansion gives
\begin{equation}
  f(x)=f(x_0)+\tfrac12\,f''(x_0)\,(x-x_0)^2+\cdots \ ,
\end{equation}
while taking $g(k)=g(\lambda x)\approx g(\lambda x_0)$ at leading order:
\begin{equation}
  S \approx \sum_{k=a}^b g(\lambda x_0)\,\exp\Bigl[\lambda f(x_0)+\tfrac{\lambda}{2}\,f''(x_0)\,(x-x_0)^2\Bigr] \ .
\end{equation}
Factoring out constants and writing the sum as a Riemann sum in $x=k/\lambda$,
\begin{equation}
  S \approx \lambda\,g(\lambda x_0)\,e^{\lambda f(x_0)}\sum_{k=a}^b\frac{1}{\lambda}\,\exp\Bigl[\tfrac{\lambda}{2}\,f''(x_0)\bigl(\tfrac{k}{\lambda}-x_0\bigr)^2\Bigr] \ ,
  \ \longrightarrow\ 
  \lambda\,g(k_0)\,e^{f(k_0)}\int_{x_a}^{x_b}e^{\frac{\lambda}{2}\,f''(x_0)(x-x_0)^2}\,\mathrm{d}x \ .
\end{equation}
Finally, by extending the integral to $\pm\infty$ (the error is exponentially small \cite{deBruijn:1981,BenderOrszag:1999}), one obtains the leading-order approximation
\begin{equation}
  S \approx g(k_0)\,e^{f(k_0)}\,\sqrt{\frac{2\pi \lambda }{-\,f''(k_0/\lambda)}} \ .
\end{equation}
Higher-order corrections arise systematically by retaining further terms in the Taylor expansions and by accounting for boundary effects when $k$ approaches the summation limits. The generalization to many dimensions and to complex-valued functions proceeds analogously to the steepest-descent extension of the Laplace method for integrals.

As an example, consider the sum (which we know exactly)
\begin{equation}
\label{eq:exactexample}
  T_n = \sum_{k=0}^n \binom{n}{k}^2 = \binom{2n}{n}\ .
\end{equation}
Using Stirling's formula
\begin{equation}
  \binom{n}{k}^2
  \sim \frac{1}{2\pi n\,x(1-x)}\,e^{n f(x)}\ ,
\end{equation}
where $x = k/n$, and 
\begin{equation}
  f(x) = -2x\log(x)-2(1-x)\log(1-x)\ .
\end{equation}
This gives
\begin{equation}
  T_n \approx \sum_{k=0}^n \frac{1}{2\pi n\,x(1-x)}\,e^{n f(x)} \ ,
  \quad \text{with} \quad x=\frac{k}{n}\ .
\end{equation}
First, we locate the saddle by solving
\begin{equation}
  f'(x)=2\log\frac{1-x}{x}=0
  \quad\to\quad x_0=\tfrac12 \ ,\quad\to\quad  f''(x_0)=-8<0 \ ,\quad f(x_0)=2\log2\ .
\end{equation}
Next, we apply the Laplace method to the sum to find
\begin{equation}
  S \approx n \frac{1}{2\pi n\,x_0(1-x_0)}\,e^{n f(x_0)}\,\sqrt{\frac{2\pi}{-n\,f''(x_0)}} = \frac{e^{n 2\log 2}}{\sqrt{\pi n}} = \frac{4^n}{\sqrt{\pi n}} \ .
\end{equation}
Finally, one can also check the approximation via Stirling's formula directly on the exact summation \eqref{eq:exactexample}
\begin{equation}
  \binom{2n}{n}^2
  \sim \frac{4^n}{\sqrt{\pi n}}\ ,
\end{equation}
in perfect agreement with the leading order of the Laplace-method approximation.

\end{appendices}

\bibliography{biblio}{}
\bibliographystyle{ieeetr}

\end{document}